\documentclass[aps,prl,twocolumn,amsmath,amssymb,superscriptaddress,showpacs,longbibliography]{revtex4-1}

\usepackage[switch]{lineno}
\usepackage{amsmath}
\usepackage{amssymb}
\usepackage{empheq}
\usepackage[parfill]{parskip}
\usepackage[active]{srcltx}
\usepackage{color}
\usepackage{array}
\usepackage{booktabs}
\usepackage{amsfonts}
\usepackage{dsfont}
\usepackage{graphicx}
\usepackage{mathtools}
\usepackage{natbib}
\usepackage{xcolor}

\begin{document}

\setlength{\parindent}{0.5cm}

\title{Time delay in the 1d swarmalator model}

\author{K. P. O'Keeffe}
 \email{kevin.p.okeeffe@gmail.com}
 \affiliation{Starling research institute, Seattle, WA 98112, USA}

 \author{Jason Hindes}
 \email{jason.m.hindes.civ@us.navy.mil}
 \affiliation{U.S. Naval Research Laboratory, Washington, DC 20375, USA}

\begin{abstract} 
We study the 1d swarmalator model augmented with time delayed coupling. Along with the familiar sync, async, and phase wave states, we find a family of unsteady states where the order parameters are time periodic, sometimes with clean oscillations, sometimes with irregular vacillations. The unsteady states are born in two ways: via a Hopf bifurcation from the phase wave, and a zero eigenvalue bifurcation from the async state. We find both of these boundary curves analytically. A surprising result is that stabilities of the async and sync states are independent of the delay $\tau$; they depend only on the coupling strength. 
\end{abstract}

\maketitle

\section{Introduction}
The 1d swarmalator model \cite{o2022collective,yoon2022sync} is a toy model for many systems which self-assemble in space and self-synchronize in time interdependently. Examples include  vinegar eels \cite{quillen2021metachronal}, tree frogs \cite{aihara2014spatio}, magnetic domain walls \cite{hrabec2018velocity}, colloidal micromotors \cite{yan2012linking,liu2021activity,zhang2020reconfigurable,bechinger2025tunable}, embryonic cells \cite{tsiairis2016self}, active spheres \cite{riedl2023synchronization}, robotic swarms \cite{barcis2019robots,barcis2020sandsbots}, and human interactions on dance floors \cite{toiviainen2025modeling}. The model consists of $N$ swarmalators, short for ``swarming oscillators", moving on a 1d periodic domain according to
\begin{align}
    \dot{x}_i &= \nu_i' + \frac{J'}{N} \sum_j \sin(x_j - x_i) \cos(\theta_j - \theta_i) \\
    \dot{\theta}_i &= \omega_i' + \frac{K'}{N} \sum_j \sin(\theta_j - \theta_i) \cos(x_j - x_i). 
\end{align}
Here $x_i, \theta_i \in S^1$ are the position and phase of the $i$-th swarmalator, $\omega_i', \nu_i'$ their natural frequencies, and $J', K'$ coupling constants. The phase dynamics model synchronization (the Kuramoto sine term) which depends on distance (the new cosine term). Conversely, the spatial dynamics model swarming/aggregation which depends on phase similarity. 

This symmetry of the 1d swarmalator model makes it one of the few models of mobile oscillators that is solvable \cite{yoon2022sync,global_sync,o2025stability}. It has facilitated exact analyses of swarmalators with disordered coupling \cite{o2022swarmalators,hao2023attractive}, phase frustration \cite{lizarraga2023synchronization}, random pinning \cite{sar2023pinning,sar2024solvable,sar2023swarmalators}, external forcing \cite{anwar2024forced,anwar2025forced}, and other effects \cite{sar2025effects,hong2023swarmalators,ghosh2025dynamics,anwar2024collective,SenthamizhanGopalChandrasekar2025Swarmalators}. See \cite{sar2025interplay} for a review of the 1d model and how it relates to 2d model.

Time-delayed coupling, however, has received less attention. That is the subject of this paper. Delays occurs in many real world systems. In engineering/robotic systems, actuation takes finite time and in chemical and biological systems, transmission through fluid media comes with unavoidable lags. Kogan et al.\ took the first step in studying delay coupled swarmalators \cite{blum2024swarmalators}. They, however, studied a 2d model, with delays confined to the phase coupling only. They found interesting physics, such as breathing disks with dynamics like aging \cite{kumpeerakij2025aging} and boundary boiling \cite{kumpeerakij2025aging}. Approximations for the breathing oscillation frequency $\omega(\tau)$ and decay rate $\gamma(\tau)$ were derived, but results about stability and bifurcations are missing; the complexity of the 2d model makes exact analyses challenging.

This paper tries to make analytic progress on delayed swarmalators by using the more tractable 1d swarmalator model. We find delay produces unsteady states not seen in the non-delayed model. We calculate explicitly how they arise: via a zero eigenvalue bifurcation from async, and via a Hopf bifurcation from the phase wave. These are the first exact results about delay coupled swarmalators and in that sense contribute to the field.

\section{Model}
We consider the 1d swarmalator with delays in both the phase and positions
\begin{align}
    \dot{x}_i &= \nu_i' + \frac{J'}{N} \sum_j \sin(x_j(t-\tau) - x_i) \cos(\theta_j(t-\tau) - \theta_i) \\
    \dot{\theta}_i &= \omega_i' + \frac{K'}{N} \sum_j \sin(\theta_j(t-\tau) - \theta_i) \cos(x_j(t-\tau) - x_i) 
\end{align}
We set $\omega'_i=\nu'_i=0$ to reduce the number of parameters to three, $(J',K',\tau)$. We also move to sum/difference coordinates $(\xi,\eta) := (x+\theta,\,x-\theta)$ which allows the model to be expressed as a pair of linearly coupled Kuramoto models,
\begin{align}
\dot{\xi}_i &= - K r_{\tau} \sin \xi_i - J s_{\tau} \sin \eta_i \label{e1}  \\
\dot{\eta}_i &= -J r_{\tau} \sin \xi_i - K s_{\tau} \sin \eta_i  \label{e2}
\end{align}
where
\begin{align}
& r e^{i \phi}= \langle e^{i \xi} \rangle \\
& s e^{i \psi}  = \langle e^{i \eta} \rangle  \\
& (J,K) = ( (J'+K')/2, (J'-K')/2)
\end{align}
The subscript $\tau$ in these denotes delayed time, e.g. $r_{\tau} := r(t-\tau)$. This clean form is what makes the model solvable. Notice when $J=0$ we recover the Kuramoto limit: the $(\xi,\eta)$ dynamics decouple into two Kuramoto models.

Now we explore the collective behavior of the model with numerical simulations and analysis.

\begin{figure}[t!]
\centering
\includegraphics[width = \columnwidth]{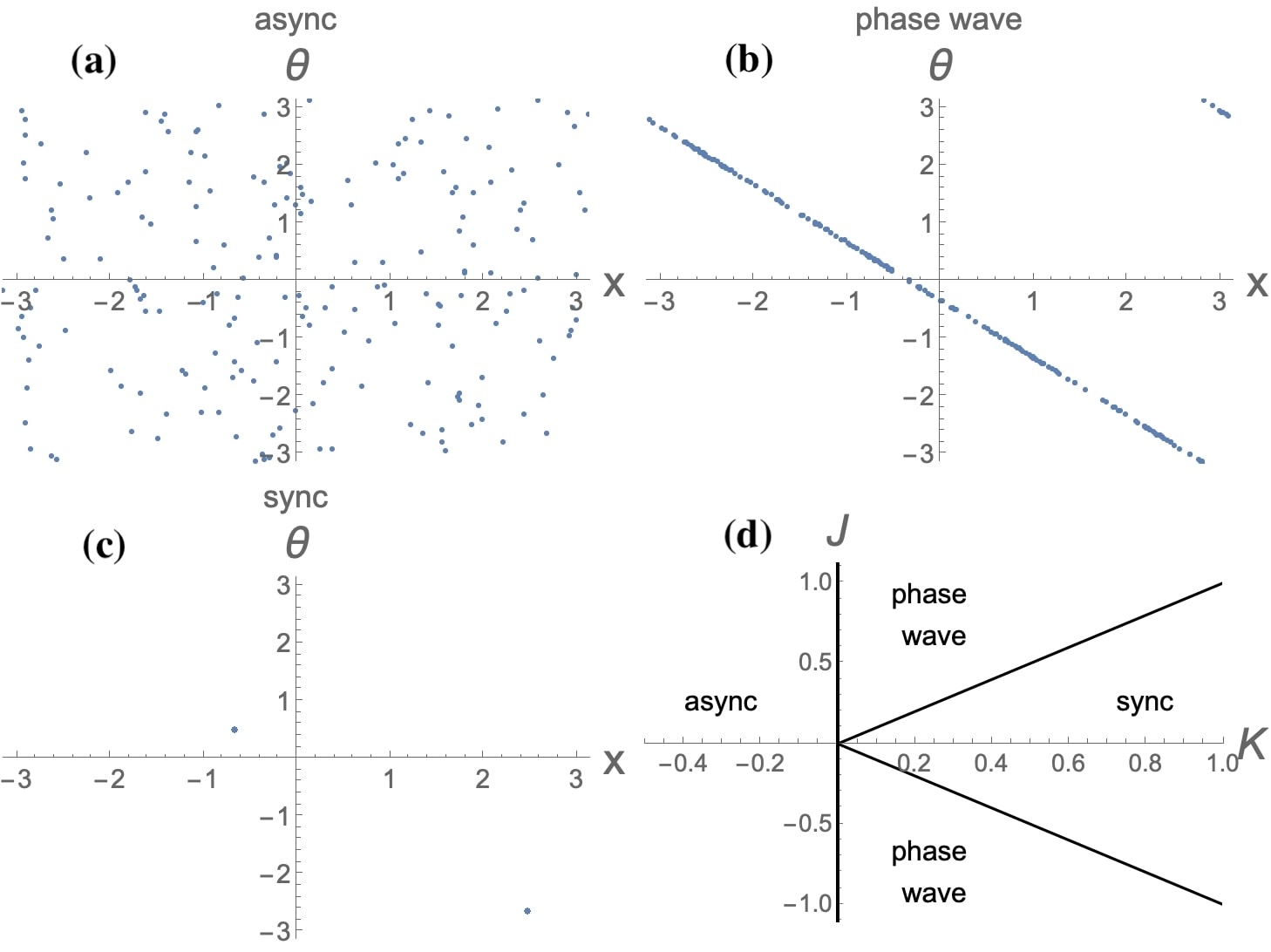}
\caption{(a)-(c) The static collective states of 1d swarmalator model in the $N \rightarrow \infty$ limit for $(J',K') = (1,-2), (1,-0.5), (1,2)$ and $\tau = 1$. Here $(dt,T,N) = (0.1,100,200)$ and initial condition were drawn uniformly at random. (d) Stability diagram in the $(J,K)$ plane for $\tau=0$, shown as baseline, to give an idea of where in parameter space the various states occur.}
\label{states-identical}
\end{figure} 

\section{Results}
\subsection{Async} 
Simulations show that for certain parameter values the asynchronous state is realized, where the swarmalators are uniformly distributed in both space and phase $\rho_0(\xi,\eta,t) = (2 \pi)^{-2}$. Here $\rho_0$ denotes the density of swarmalators and is interpreted in the Eulerian sense. Observed in previous studies, this async state is sometimes called the $(r,s) = (0,0)$ state \cite{yoon2022sync}, since both order parameters are zero in the state. Figure~\ref{states-identical}(a) plots the async state as a scatter plot in the $(x, \theta)$ plane.

\textit{Stability}. We can find the threshold at which async loses stability by plugging a linear perturbation
\begin{align}
    \rho(\xi,\eta,t) = (2 \pi)^{-2} + \epsilon \rho_1(\xi, \eta, t)
\end{align}
into the continuity equation,
\begin{equation}
    \dot{\rho} + \nabla \cdot ( v \rho ) = 0 \label{cont}
\end{equation}
where the velocities $v=(v_\xi,v_\eta)$ are the right--hand sides of Eqs.~\eqref{e1},~\eqref{e2} expressed in $(\xi,\eta)$:
\begin{align}
    v_\xi(\xi,\eta,t) &= - K\, r_\tau \sin \xi \;-\; J\, s_\tau \sin \eta, \label{vx-lin}\\
    v_\eta(\xi,\eta,t) &= - J\, r_\tau \sin \xi \;-\; K\, s_\tau \sin \eta, \label{vy-lin}
\end{align}
with the delayed order parameters
\begin{align}
    r_\tau e^{i\phi_\tau} := \langle e^{i\xi}\rangle_{t-\tau}, \qquad
    s_\tau e^{i\psi_\tau} := \langle e^{i\eta}\rangle_{t-\tau}.
\end{align}
In the async base state $r=s=0$, so $v_0\equiv 0$ and the linearized velocity is
\begin{align}
    & v = v_0 + \epsilon v_1 = \epsilon v_1 \\ 
    & v_1 = \big(- K\, r_\tau \sin \xi - J\, s_\tau \sin \eta,\; - J\, r_\tau \sin \xi - K\, s_\tau \sin \eta\big). \label{v1}
\end{align}
Three steps are needed from here. First, observe that normalization requires the integral of the perturbed density $\rho_1$ to be zero
\begin{align}
    \int \rho_1(\xi, \eta, t) \, d \xi \, d \eta = 0.
\end{align}
Second, observe that the decomposition $\rho = \rho_0 + \epsilon \rho_1$ induces the same decomposition in the velocity $v = v_0 + \epsilon v_1 = \epsilon v_1$, since $v_0=0$ in async; the $v_1$ is given in \eqref{v1}. Third, we expand the perturbation $\rho_1$ in a Fourier series
\begin{align}
    \rho_1(\xi,\eta,t) 
    &= \frac{1}{(2\pi)^2}\!\!\sum_{m,n\in\mathbb{Z}} a_{mn}(t)\, e^{i m \xi + i n \eta}
\end{align}
with $a_{00} = 0$, which follows from the normalization. Notice that the perturbed order parameters are related to $a_{10}, a_{01}$
\begin{align}
    r_\tau &= \langle e^{i\xi}\rho_1\rangle_{t-\tau} = a_{-1,0}(t-\tau) \\
    s_\tau &= \langle e^{i\eta}\rho_1\rangle_{t-\tau} = a_{0,-1}(t-\tau).
\end{align}
Projecting \eqref{cont} with $v$ from \eqref{v1} onto the $(m,n)=(1,0)$ and $(0,1)$ harmonics (using $\nabla\!\cdot(v\rho_0)=\rho_0(\partial_\xi v_\xi + \partial_\eta v_\eta)$) yields the closed linear system
\begin{align}
    \dot a_{10}(t) &= \frac{K}{2}\, a_{10}(t-\tau), \label{a10-eq}\\
    \dot a_{01}(t) &= \frac{K}{2}\, a_{01}(t-\tau), \label{a01-eq}\\
    \dot a_{mn}(t) &= 0, \qquad \text{for all other }(m,n)\neq(1,0),(0,1).
\end{align}
Only the first modes are dynamically active. This convenient property stems from the fact that the 1d swarmalator model is composed of sine terms in $\xi, \eta$ that are linearly coupled; that is what makes the model solvable. Searching for normal modes $a_{10}(t)=b\,e^{\lambda t}$ (and identically for $a_{01}$) gives the characteristic equation
\begin{align}
    \lambda\, e^{\lambda \tau} = \frac{K}{2}. \label{char-async}
\end{align}
This contains all the information we need. First notice that when $\tau = 0$, \eqref{char-async} collapses to $ \lambda = K/2$. This implies the critical coupling is $K_c = 0$, consistent with previous works \cite{o2022collective}. For non-zero $\tau$ the $K_c$ actually stays the same. To see this, write \eqref{char-async} as
\begin{align}
    \lambda = \frac{1}{\tau}\, W_k\!\Big(\frac{K\tau}{2}\Big), \qquad k\in\mathbb{Z},
\end{align}
in terms of the Lambert $W$ function. Then $\Re\,\lambda>0$ iff $K>0$ on the principal branch $W_0$, whereas for $K\le 0$ all branches satisfy $\Re\,\lambda\le 0$. Our final result is that async is stable for any values of delay $\tau$ and coupling $J$, as long as $K$ is smaller than the critical value
\begin{align}
\Aboxed{
 K_c = 0.
}
\end{align}
Delay, it turns out, does not affect the stability of the async state at all.

\subsection{Phase wave}
Delay does, however, affect another of the model's collective states: the phase wave, where the swarmalators' phases and positions are correlated perfectly $x_i \pm \theta_i + const \Longleftrightarrow (\xi_i, \eta_i) = (const, 2 \pi i / N), (2 \pi i / N, const) $. Figure~\ref{states-identical}(b) plots a clockwise wave (counter clockwise waves are realized with equal frequency). Phase waves mimic several real world swarmalator states such as the metachronal waves of vinegar eels \cite{quillen2021metachronal}, vortex arrays of sperm \cite{riedel2005self}, and asters in magnetic colloids \cite{snezhko2011magnetic}.

\textit{Stability}. The approach is the same as before: linearize around equilibrium in density space. We analyze, without loss of generality, a $\xi$-wave with zero offset $const=0$,
\begin{align}
    \rho_0(\xi,\eta) = \frac{1}{2\pi}\delta(\xi)
\end{align}
Notice the density now has a $\delta$-function. This makes the calculation much richer than that of async. It is the main analytic contribution of this paper. 

Letting $\rho = \rho_0 + \epsilon \rho_1(\xi,\eta,t)$ and expanding $\rho_1$ in a Fourier series leads to the following eigenvalue equation for the principal eigenmode $b(\xi,\eta)$
\begin{align}
   \lambda b &= K b \cos \eta + \sin \eta \left( K b_{\eta}  + J b_{\xi} \right) - \frac{s_1}{2 \pi} K \sin \phi_1 \delta'(\eta) \nonumber \\
   &+\frac{r_1}{2 \pi} \Big( K \delta(\eta) \cos(\xi - \phi_1) + J \sin(\xi - \phi_1) \delta'(\eta) \Big) \label{ee} 
\end{align}
Notice in Eq.~\eqref{ee}, there are delta functions and their derivatives on the rhs in the $\eta$ direction. This means the eigenfunction $b$ must be a distribution in $\eta$, but a regular function in $\xi$. More formally, $b$ must be in the space of tempered distributions defined on the unit torus $b \in S(\mathbb{T}^1)$. Actually, we take a subset of this space for which the first integral is zero $\int b = 0 $, which we define as $S'(\mathbb{T}^1)$. The benefit of working with this function space is that \textit{any} distribution $f(x)$ with support at $x=0$ can be represented as a finite sum of delta functions and their derivatives $f(x) = \sum_{n} \delta^{(n)}(x)$ \cite{friedlander1998introduction}. A \textit{general} representation for our eigenvector $b$ is thus
\begin{align}
    b(\xi,\eta) = \frac{1}{2\pi} \Big( a_0(\eta) + \cos(\xi) a_1(\eta) + \sin(\xi) b_1(\eta) \Big) + h.o.t.
\end{align}
where $h.o.t$ means the higher Fourier harmonics (they are dynamically unimportant, just as in the async calculation in the previous section). The novelty is that the Fourier coefficients are \textit{delta series}
\begin{align}
    & a_0(\eta) = \sum_{n=1} c_{n} \delta^{(n)}(\eta) \label{qa2} \\
    & a_1(\eta) = \sum_{n=0} d_{n} \delta^{(n)}(\eta) \label{qa3} \\
    & b_1(\eta) = \sum_{n=0} e_{n} \delta^{(n)}(\eta) \label{qa4}
\end{align}
Notice the sum for $a_0$ starts at $n=1$ but those for $a_1, b_1$ start at $n=0$, which comes from the normalization condition. Plugging in this ansatz and collecting coefficients leads to a set of simultaneous equations for $c_n, d_n, e_n$ and $\lambda$. One set exists for the zeroth Fourier mode, and a symmetric pair for the first two eigenmodes in $\xi,\eta$. The $\lambda$ from the first modes are the critical ones and satisfy the following characteristic equation
\begin{align}
\left(J^2-K^2\right) e^{-\lambda  \tau }+2 \lambda ^2+\lambda  K \left(2-e^{-\lambda  \tau }\right) = 0. \label{f1}
\end{align}
In the limit $\tau = 0$, this reduces to $J^2 - K^2 + 2 \lambda^2 + \lambda K = 0 $ which is consistent with previous works \cite{o2022collective}. 

Now we must solve Eq.~\eqref{f1}. How do we do so? We look for a Hopf bifurcation by setting $\lambda = i\omega$ ($\omega>0$) and write $e^{-i\omega\tau}=\cos(\omega\tau)-i\sin(\omega\tau)$. 
Equation~\eqref{f1} can be rearranged as
\[
e^{-\lambda\tau}\big[(J^2-K^2)-K\lambda\big]+2\lambda(\lambda+K)=0.
\]
With $\lambda=i\omega$, separating real/imaginary parts gives the linear system for 
$c:=\cos(\omega\tau)$ and $s:=\sin(\omega\tau)$:
\begin{align}
(J^2-K^2)\,c - (K\omega)\,s &= 2\omega^2, \label{eq:R}\\
(K\omega)\,c + (J^2-K^2)\,s &= 2K\omega. \label{eq:I}
\end{align}
Solving \eqref{eq:R}--\eqref{eq:I} for $(c,s)$ yields
\begin{align}
\cos(\omega\tau) &= \frac{2J^2\,\omega^2}{(J^2-K^2)^2+K^2\omega^2} \\
\sin(\omega\tau) &= \frac{2K\omega\,(J^2-K^2-\omega^2)}{(J^2-K^2)^2+K^2\omega^2}. \label{eq:cs}
\end{align}
\begin{figure}[t!]
\centering
\includegraphics[width = \columnwidth]{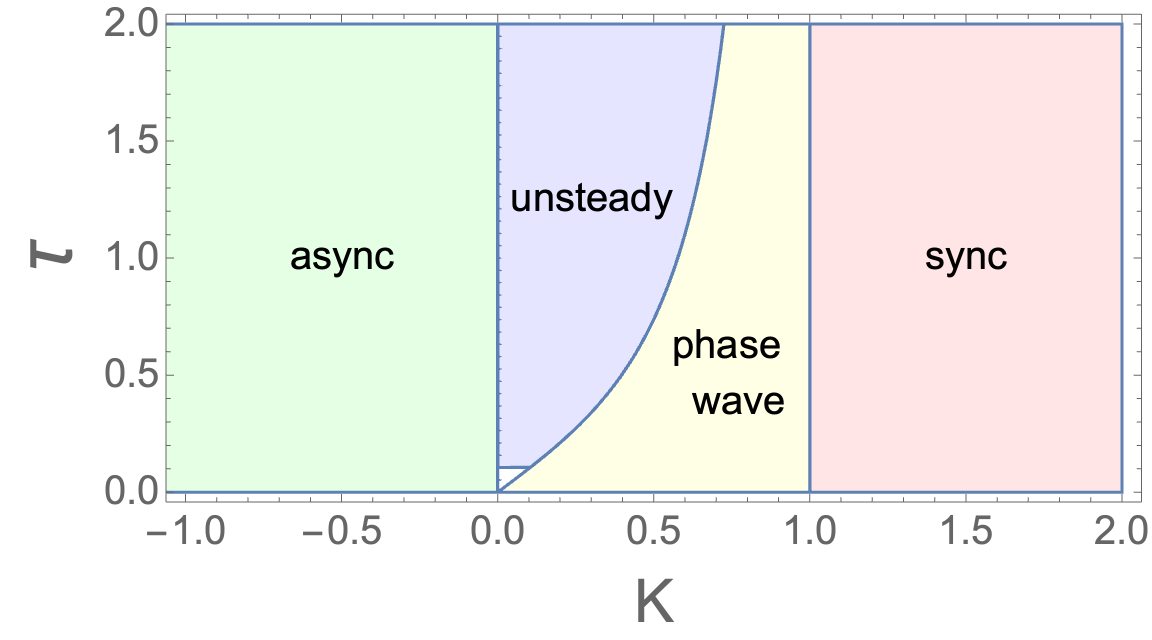}
\caption{Bifurcation diagram in the $(K,\tau)$ plane for $J = 1$. }
\label{bif-diagram-1}
\end{figure}
Imposing $\cos^2(\omega\tau)+\sin^2(\omega\tau)=1$ gives a quartic for $\omega$ that is quadratic in $\omega^2$:
\[
4\omega^4+3K^2\omega^2-(J^2-K^2)^2=0,
\]
so the admissible (positive) solution is
\begin{align}
\omega_c^2 \;=\; \frac{-3K^2+\sqrt{\,16J^4-32J^2K^2+25K^4\,}}{8}. \label{eq:omega}
\end{align}
Finally, from \eqref{eq:cs},
\begin{align}
\cos(\omega_c\tau_c) &= \frac{2J^2\,\omega_c^2}{(J^2-K^2)^2+K^2\omega_c^2} \\
 &= \frac{4J^2}{\sqrt{\,16J^4-32J^2K^2+25K^4\,}+5K^2},
\end{align}
which implies
\begin{align}
\Aboxed{%
\tau_c &= \frac{2 \sqrt{2} \sec^{-1}\!\left(\frac{\sqrt{-32 J^2 K^2+16 J^4+25 K^4}+5 K^2}{4 J^2}\right)}
{\sqrt{\sqrt{-32 J^2 K^2+16 J^4+25 K^4}-3 K^2}}%
} \label{eq:tau}
\end{align}
Eq.~\eqref{eq:tau} is our final answer, the stability threshold we were looking for. 

Figure~\ref{bif-diagram-1} plots the critical delay $\tau_c$ in the $(K,\tau)$ plane for $J=1$. Notice as $\tau \rightarrow 0$ it merges with the async threshold $K = 0$. On the other extreme, as $\tau \rightarrow \infty$ it merges with $K = J = 1$, which is the stability border for the sync state, which we analyze next. The unsteady state that is born via Hopf we analyze after that. 
\begin{figure}[t!]
\centering
\includegraphics[width = \columnwidth]{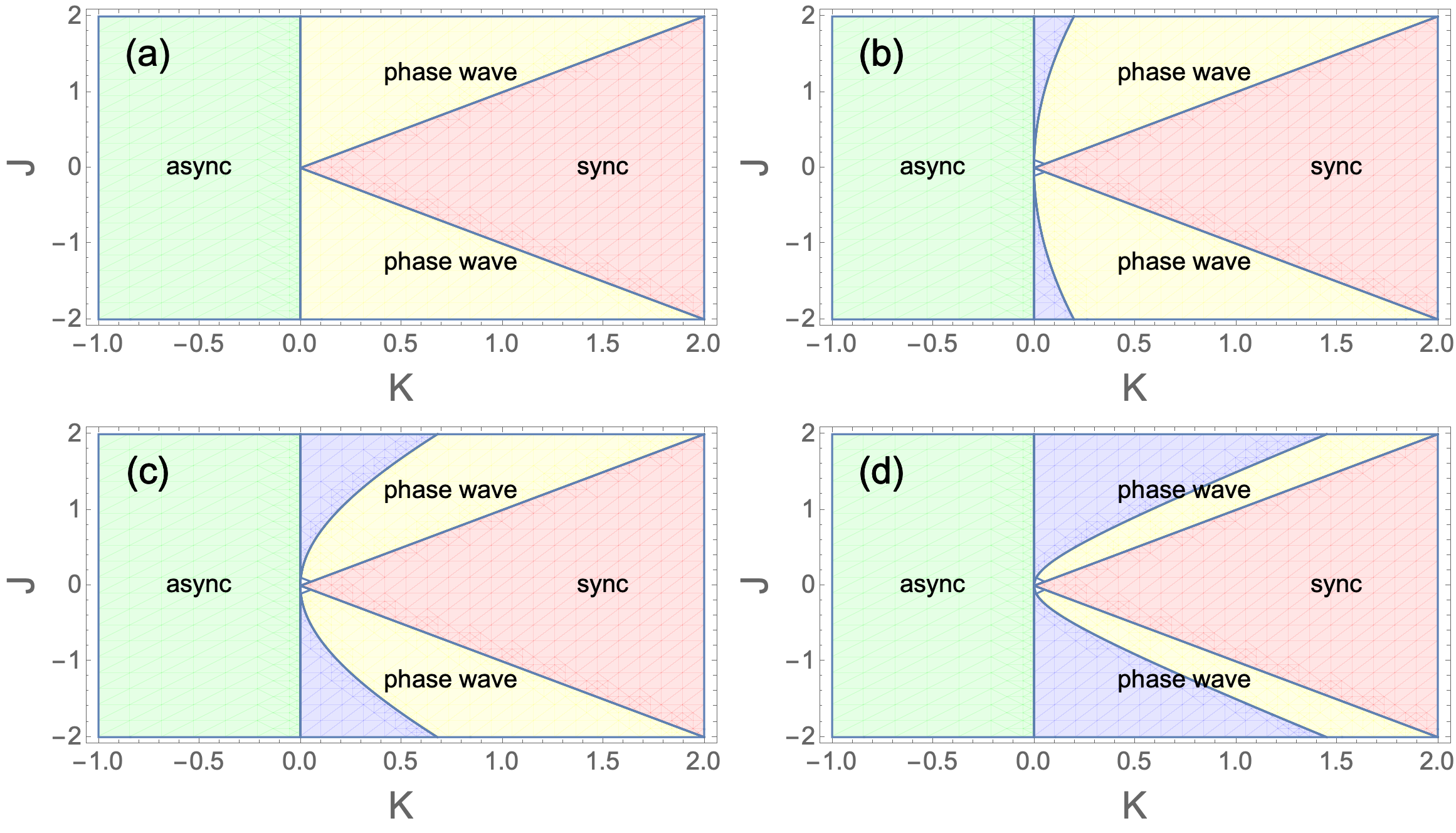}
\caption{Bifurcation diagram in the $(J,K)$ plane for $\tau = 0, 0.1, 0.25, 1.5$. The unsteady region is plotted in blue and is unlabeled due to space constraints.}
\label{bif-diagram-2}
\end{figure}

\subsection{Sync}
The sync state is defined by all swarmalators having identical coordinates $\xi_i, \eta_i = const, const$; equivalently, $r=s=1$, plotted in Figure~\ref{states-identical}(c). To find its stability, consider the equilibrium density
\begin{align}
    \rho_0(\xi,\eta) = \delta(\xi) \delta(\eta)
\end{align}
where we set the offset constants to be zero wlog. A linearization calculation similar to that of async yields
\begin{align}
(J \pm K) \left(1-e^{-\lambda  \tau }\right)+\lambda = 0 \label{g1}
\end{align}
To solve these, first set $a=J \pm K $. Then
\[
a\!\left(1-e^{-\lambda\tau}\right)+\lambda=0
\;\Longleftrightarrow\;
e^{-\lambda\tau}=1+\frac{\lambda}{a}.
\]
Let \(y:=1+\lambda/a\). Then \(e^{-a\tau(y-1)}=y\), i.e.
\[
(a\tau y)\,e^{a\tau y}=a\tau e^{a\tau}
\quad\Rightarrow\quad
a\tau y=W_k\!\big(a\tau e^{a\tau}\big),
\]
so the spectrum on each branch \(k\in\mathbb{Z}\) is
\begin{equation}
\lambda_k^{(\pm)}(\tau)=\frac{1}{\tau}\,W_k\!\Big((J\!\pm\!K)\,\tau\,e^{(J\pm K)\tau}\Big)\;-\;(J\!\pm\!K).
\label{eq:lambert-sync}
\end{equation}
A Hopf crossing would require \(\lambda=i\omega\) with \(\omega>0\). Writing
\(e^{-i\omega\tau}=\cos(\omega\tau)-i\sin(\omega\tau)\) in
\(a(1-e^{-i\omega\tau})+i\omega=0\) and separating real/imaginary parts gives
\[
a\bigl[1-\cos(\omega\tau)\bigr]=0,\qquad \omega+a\sin(\omega\tau)=0.
\]
Thus either \(a=0\) or \(\cos(\omega\tau)=1\Rightarrow \sin(\omega\tau)=0\Rightarrow \omega=0\).
Hence no genuine Hopf bifurcation can originate from \eqref{g1}. The only way
these blocks can become marginal is when
\begin{align}
\Aboxed{
K_c = \pm J
}
\end{align}
Like the async state, the sync threshold $K_c=\pm J$ is independent of $\tau$.

\textit{Stability finite-$N$}. We can also find the stability of the sync state for finite $N$. For a delay differential equation of form
\begin{align}
    \dot{z} = f(z, z(t-\tau))
\end{align}
where $z \in \mathbb{R}^N$, the characteristic equation is
\begin{align}
    \det \Big[ \lambda I - M_0 + e^{\lambda \tau} M_1 \Big] = 0
\end{align}
Here $M_0$ is $\partial_{z}$ and $M_1$ is $\partial_{z(t-\tau)}$. For our problem,
\[
M_0 = \begin{pmatrix}
- \frac{K (N - 1)}{N} I_N & - \frac{J (N - 1)}{N} I_N \\
- \frac{J (N - 1)}{N} I_N & - \frac{K (N - 1)}{N} I_N
\end{pmatrix}
\]
\[
M_1 = \begin{pmatrix}
\frac{K}{N} B & \frac{J}{N} B \\
\frac{J}{N} B & \frac{K}{N} B
\end{pmatrix}
\]
with \( B = A - I_N \), and \( A \) the \( N \times N \) all-ones matrix.

After some algebra we find 
\begin{align}
    \lambda_N \pm \frac{N-1}{N} (J \pm K) \Big( 1 -  e^{\lambda \tau} \Big) = 0 
\end{align}
A similar calculation to the continuum case shows $K_c = \pm J$; the stability threshold for the finite-$N$ matches the continuum limit $N \rightarrow \infty$.

\subsection{Unsteady states}
\begin{figure}[t!]
\centering
\includegraphics[width = \columnwidth]{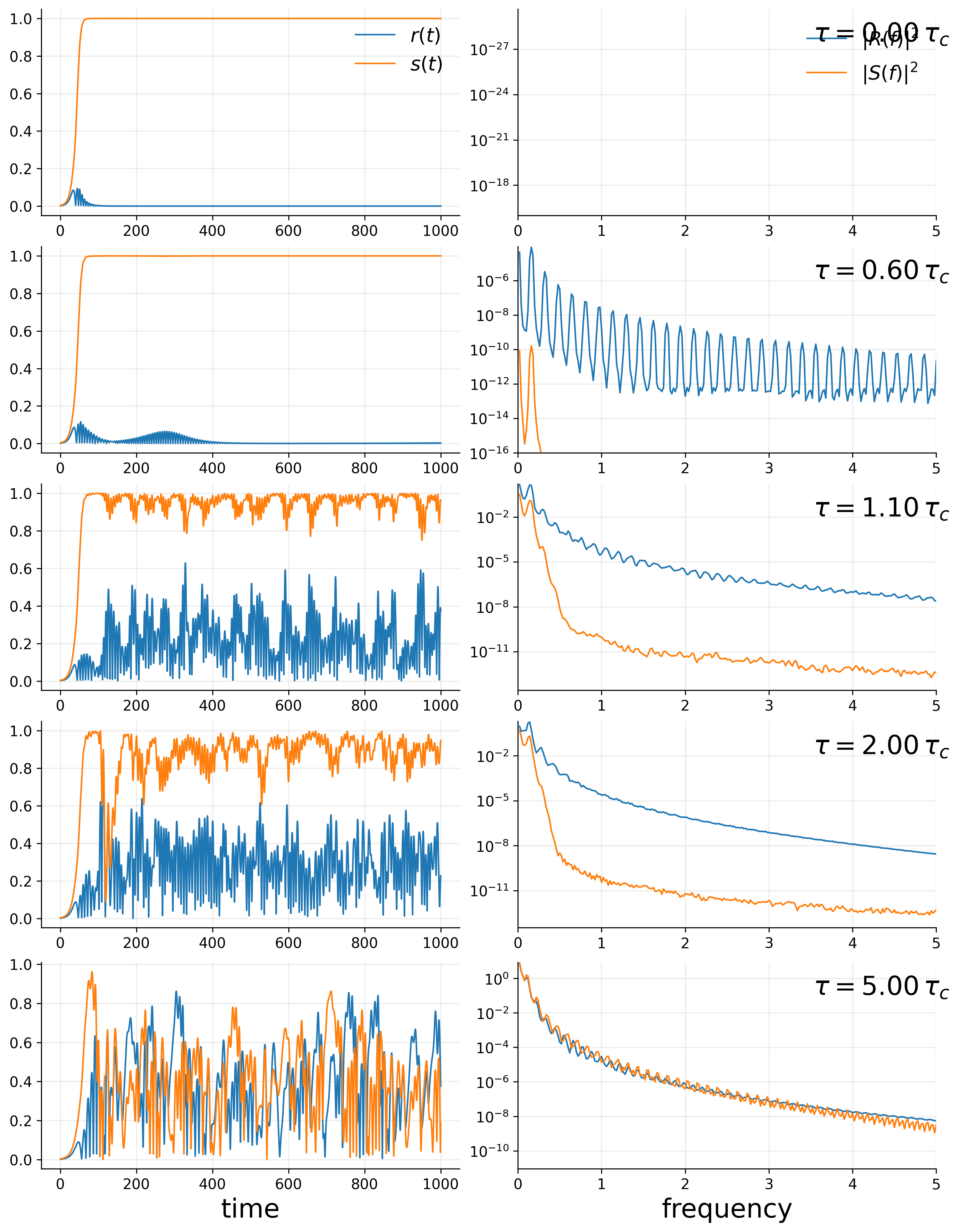}
\caption{Left column, time series of order parameters for different values of $\tau$. Right column, associated power spectra. Parameters were $(J',K') = (1,-0.5)$ and simulations were for $N= 10^5$ for $T=10^3$ time units with stepsize $dt=0.01$.}
\label{time-series}
\end{figure}

\begin{figure}[t!]
\centering
\includegraphics[width = \columnwidth]{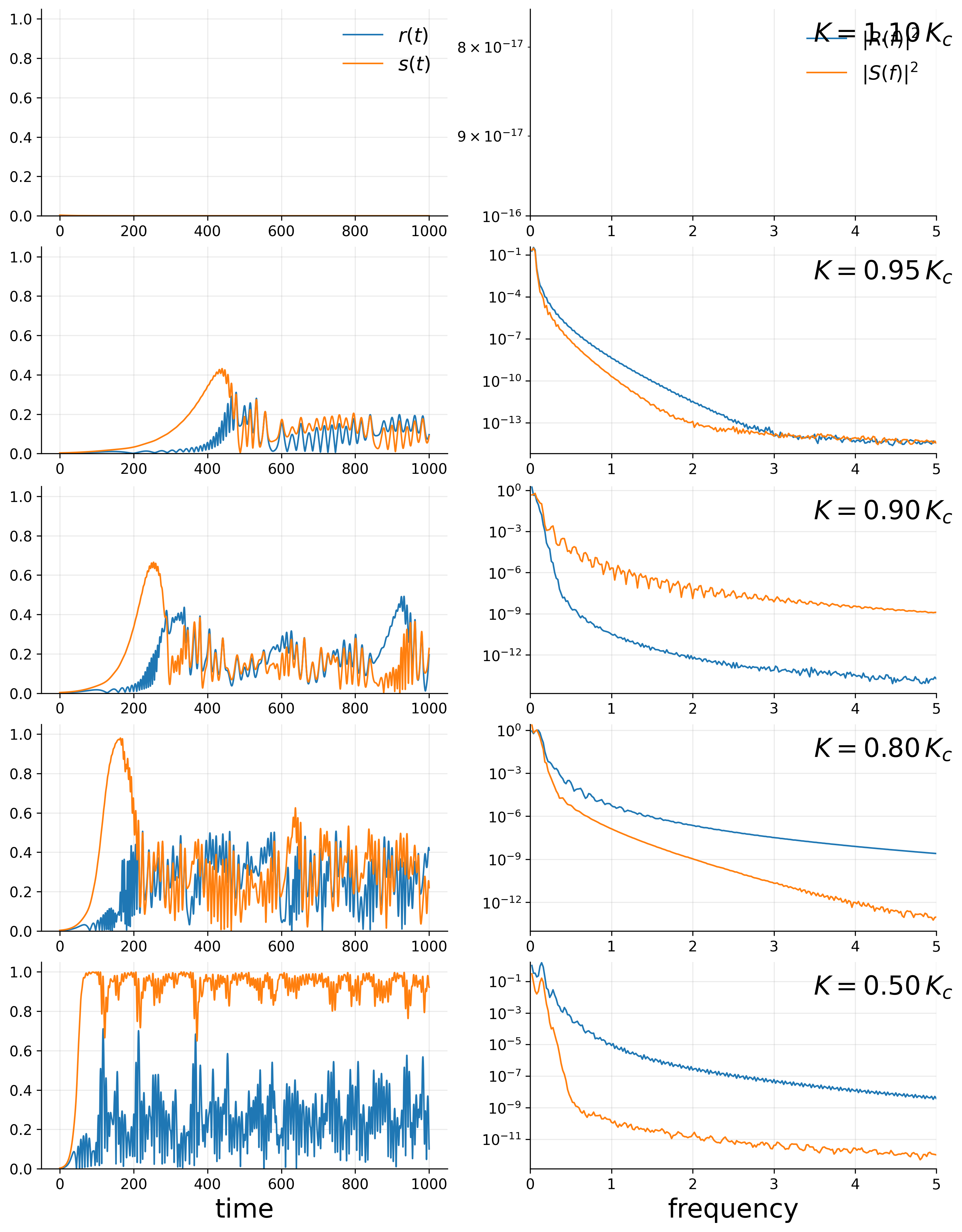}
\caption{Same as Figure~\ref{time-series} except we vary $K$ and keep $\tau=1$ constant.}
\label{time-series-sweep-K}
\end{figure}

Let's distill our results so far. We found the stability curves for the model's three static states: async, phase wave, and sync. Surprisingly, the stability of async and sync do not depend on the delay $\tau$. The phase wave, on the other hand, destabilizes via a Hopf bifurcation that depends on all three parameters $\tau, J, K$. Figure~\ref{bif-diagram-1} summarizes these findings with the bifurcation diagram in $(K,\tau)$ space. For completion's sake, Figure~\ref{bif-diagram-2} shows the bifurcation diagram in $(J,K)$ space for different values of $\tau$; the takeaway is that as $\tau$ gets larger, it occupies more and more of the phase wave's stability region.

Now we turn from statics to dynamics. We study the unsteady state that bifurcates from the phase wave. Figure~\ref{time-series} shows time series of the order parameters (left column) and their associated power spectra (right column) as we tune $\tau$ past the Hopf point. A transition from simple, oscillatory motion to wilder, vacillatory motion is clear in both the time and frequency domain. 

Below the Hopf point ringing is observed: $s(t)$ shows small, regular oscillations and the PSD has a single sharp peak with weak harmonics. Above the Hopf point, the oscillations grow and become amplitude--modulated; the dominant frequency drifts lower and the spectrum broadens with enhanced low--$f$ power. For large delay ($5.00\,\tau_c$), the time series becomes intermittent/bursty and the PSD is broadband and low--frequency--dominated.

The broad power spectra for large $\tau$ indicate the motion may be chaotic. To test this, we employed the 0-1 test for chaos, a binary test that distinguishes between regular ($\mathcal{K} \approx 0$) and chaotic ($\mathcal{K} \approx 1$) dynamics from time series data. We applied the test to $r(t)$ and $s(t)$ in a range of system sizes and sampling resolutions to rule out finite-size effects and numerical artifacts. In all cases, we found $\mathcal{K} < 0.05$. The motion is thus likely quasiperiodic, not chaotic.

Casting an eye back to the bifurcation diagram in Figure~\ref{bif-diagram-1}, we see the unsteady states may also bifurcate from the async state. Recall our analysis showed this was not a Hopf, rather a zero $\lambda$ bifurcation. Figure~\ref{time-series-sweep-K} explores this transition by again plotting time series and their spectra, but this time freezing $\tau=1$ and sweep $K$ past its critical point. For $K>K_c$, both $r(t), s(t)$ hit zero quickly, as expected in the async state. But for $K<K_c$, small oscillations are born. Their amplitudes get larger and larger and eventually resemble the time series from Figure~\ref{time-series}.

\subsection{Discussion}
\begin{figure}[t!]
\centering
\includegraphics[width = \columnwidth]{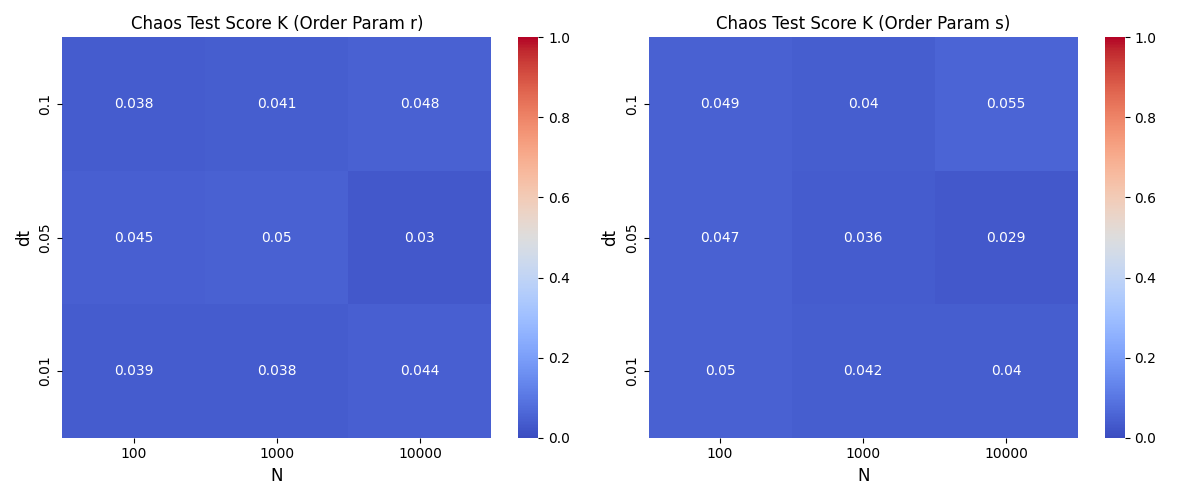}
\caption{0-1 test for chaos in the unsteady regime ($\tau = 5.0\tau_c$). The test score $K \approx 0$ across varying system sizes $N$ and time steps $dt$ indicates the dynamics are non-chaotic.}
\label{chaos}
\end{figure}
We have presented the first theoretical study of delay coupled swarmalators with exact results. We derived the stability boundaries of the model's three static collective states which allowed us to draw the full phase diagram the $(J,K,\tau)$ space. We also found a family of unsteady states not seen in non-delayed systems. These dynamics might be observable in populations of vinegar eels, sperms, or other real world swarmalators. 

A key technical contribution is the distributional perturbation analysis around the phase wave (using delta-function series in one coordinate), which yields exact Hopf thresholds despite the infinite-dimensional nature of delayed systems. This technique extends straightforwardly to higher-dimensional swarmalator models and to cases with parametric disorder or additional degrees of freedom---offering a promising route to tackle several currently open problems in the field \cite{sar2026interplay}.

One implication of our findings is that any fluctuations observed in empirical time series of $r(t), s(t)$ may be due to delayed communications, and not due to thermal effects, finite $N$ effects, or heterogeneities in natural frequencies of couplings. A real world system where this might be relevant is groups of sperm confined to 1d copper rings \cite{creppy2016symmetry}, whose interactions are mediated hydrodynamically and thus are non-instantaneous.

Future work could extend our study to the regime where $\omega', \nu'$ are non-zero (recall we set these to zero to reduce the number of free parameters to three). One could also extend the model to 2 or 3 spatial dimensions \cite{o2024solvable, anwar2025forced} or add external forcing. Finally, one could derive some \textit{global} stability results \cite{global_sync} which may be useful in swarm robotics \cite{schilcher2025multicircular} in which convergence results in the presence of delays are useful. 

    

%

\end{document}